\documentstyle[12pt,epsfig]{article}
\begin{document}

\bigskip

\centerline {\bf {Stability of Accretion Disc around a Black Hole: effects of}}
\centerline {\bf {Bremsstrahlung and Synchrotron Cooling}}
\medskip
\bigskip
\bigskip
\centerline {Sivakumar G. Manickam}
\centerline {Harish-Chandra Research Institute,}
\centerline {Chhatnag Road, Jhunsi, Allahabad 211019, India}
\centerline {email: \it{sivman@mri.ernet.in}}

\medskip
\bigskip
\bigskip

\noindent {\bf Abstract}

\noindent We perform stability analysis of an accretion disc around a stellar mass
black hole. The cooling of the accretion flow due to bremsstrahlung and synchrotron
radiation processes are considered. The solutions with shock are perturbed at the shock location
and the radial dependence of perturbation is obtained by solving the equations numerically
using fourth-order Runge-Kutta method. The perturbations are assumed to be infinitesimal,
thereby allowing a linear analysis. The frequencies that solve the equations are found
and the existence of unstable frequencies implies instability. 

\noindent Running Title : Accretion disc stability - Bremsstrahlung and Synchrotron cooling 

\bigskip


\ \\
\noindent {\bf 1. Introduction}
 
\noindent Most of the popular studies in accretion discs consider steady-state
situations (Bondi 1952; Shakura \& Sunyaev 1973;
Novikov \& Thorne 1973;
Narayan \& Yi 1994; Paczy{\'n}ski 1998) i.e. where things are not
changing with time.
Steady state approximation is applicable only when the time scale
for change of the physical phenomena is large. We will be better equipped to explain
the transient and time-dependent phenomena by relaxing the assumption
that the flow is steady in our analysis.
The steady state properties of discs are mostly independent of viscosity and hence their observations
do not lead to the quantification of the viscosity.
But the time-dependent behaviour of the disc is more sensitive to the viscosity mechanism
operating in the disc. Quantitative information of the viscosity
can be derived by observations of the variable phenomena. An example of the time-dependent
phenomena is the outburst of a dwarf novae.

Analysing the stability of a fluid is one of the most difficult problems to handle in fluid dynamics.
The monograph of Chandrasekhar (1961) uses variational principle which is an approximate method,
to tackle the problem of hydrodynamic and hydromagnetic stability analytically.
Langer, Chanmugam \& Shaviv (1982, hereafter LCS82) consider column accretion in the case of a
white dwarf in the presence of optically thin bremsstrahlung cooling. When the magnetic field is strong
the matter is accreted along the field
lines onto the polar caps. The energy is released in form of extreme ultraviolet, soft X-ray and hard
X-ray radiation. The time-dependent solution was obtained by a numerical simulation which uses
finite difference scheme on an Eulerian grid. The post shock flow was found to be thermally unstable
and the shock undergoes periodic oscillations. This oscillation was unexpected and interesting and
the properties of the oscillation are dependent on the accretion rate and mass and radius of the white
dwarf. The time scale of oscillation corresponds to post-shock cooling time scale.
In Chanmugam, Langer \& Shaviv (1985, hereafter CLS85) cooling due to cyclotron process also was included.
The presence of cyclotron cooling tends to stabilize the flow if the
magnetic field is more than $2 \times 10^7$ gauss.
Balbus \& Hawley (1991, hereafter BH91) consider magneto-hydrodynamics of accretion. The flow was
found to be unstable
to axisymmetric disturbances in the presence of a weak magnetic field.
The instability is
local and extremely powerful.
The fluid motions create poloidal and toroidal magnetic field components.
Chandrasekhar and Velikhov studied a related problem of Couette flow. 
This is a  promising mechanism for generating
viscosity but the derived viscosity coefficient value is not sufficient to explain the deduced
accretion rates.

     There are different time scales in which the disc structure can vary. They are the viscous time scale
$t_{visc}$, dynamical time scale $t_\phi$,
time scale in which transverse hydrostatic equilibrium is established $t_z$,
thermal time scale $t_{th}$ and cooling time scale $t_{cool}$.

$t_{visc} \sim {R^2 \over \nu} \sim {R \over v_R}$ is the time scale for matter in the disc to diffuse due to viscous
\indent \indent \ \ \ \ \ torques

$t_\phi = {R \over v_\phi}$ is the time scale of rotational motion

$t_z = {H \over c_s}$ is the time scale in which perturbations along z directions are smoothed
\indent \indent \ \ out

$t_{th}$ is the time scale for thermal perturbation to re-adjust

$t_{cool}$ is the cooling time scale in which thermal energy is dissipated due to various 
\indent \ \ \ \ \ \ cooling processes

Oscillating shocks may have significance for accretion onto compact objects as it
could potentially explain the observed X-ray variabilities, and in particular quasi-periodic
oscillations (LCS82; Molteni, Sponholz \& Chakrabarti 1996; Chakrabarti \& Manickam 2000).
We perform global stability analysis for an axisymmetric system and hence the perturbations
take the form $f_1(r) e^{i(kz+m\phi-\omega t)}$, where $f_1(r)$ is an infinitesimal and it can
be any one of the flow variables. 

\ \\
\noindent {\bf 2. Time-dependent infinitesimal perturbation from steady state}

\noindent We assume that the time-dependent solution can be expressed as a sum of time-independent solution and a
time-dependent infinitesimal. Any arbitrary time-dependent infinitesimal can be expressed as sum
of normal modes. We choose to express them as sines and cosines. The complex representation of
the Fourier
series is used to study both the growth and decay of the modes.
The variables are expressed as,

$$
\rho = \rho_0(r) + \rho_1(r) e^{i(kz+m\phi-\omega t)},
$$

$$
v_r = v_{r_0}(r) + v_{r_1}(r) e^{i(kz+m\phi-\omega t)},
$$

$$
v_\phi = v_{\phi_0}(r) + v_{\phi_1}(r) e^{i(kz+m\phi-\omega t)},
$$

$$
v_z = v_{z_1}(r) e^{i(kz+m\phi-\omega t)},
$$

\noindent and,

$$
p = p_0(r) + p_1(r) e^{i(kz+m\phi-\omega t)}.
$$

\noindent The unperturbed quantities have subscript 0 and infinitesimals have
subscript 1.
The perturbations $\rho_1, v_{r_1}, v_{\phi_1}, v_{z_1}, p_1$ are complex quantities.
The wave number $k$ takes real values, m is an integer (since $\phi+2\pi$ should be same as $\phi$)
and the frequency
$\omega$ $(=\omega_R+i\omega_I)$ is a complex number.
The presence of positive imaginary frequency ($\omega_I$) signals instability.

\ \\
\noindent {\bf 3. Linear stability analysis}

\noindent The local stability analysis may miss an unstable mode which can be seen only in a global analysis.
In the global analysis that is performed here, we assume that the
perturbations are infinitesimals. So it is possible to neglect the higher order terms of
perturbations other than the first order. This permits us to perform a linear analysis.
For a non-linear
analysis one has to rely on computer simulations.
The surface $r=r_s$ where the Mach number reaches unity can be considered as a sound
horizon, as the perturbations generated within that surface can be advected only inwards and it cannot propagate out.

We use pseudo-Newtonian description of a Schwarzschild black hole with the help of a potential,
prescribed by Paczy{\'n}ski \& Wiita (1980).
This pseudo-potential produces the positions of marginally stable and marginally bound
orbits correctly and the efficiency approximately same as that of the effective potential
of a Schwarzschild black hole.
The equations describing the dynamics of the flow are the conservation
equations of mass, momentum and energy. The variables velocity, density and pressure are
denoted as ${\bf v}(v_r, v_{\phi}, v_z), \rho$  and $p$ respectively. Cylindrical coordinate
system $(r, \phi, z)$ is used and the time-dependent equations take the form,

\noindent Continuity equation:

$$
{{\partial \rho}\over{\partial t}} + {1 \over r}{\partial \over \partial r}(r \rho v_r)
+ {1 \over r}{\partial \over \partial \phi}(\rho v_\phi) + {\partial \over \partial z}
(\rho v_z) = 0,
\eqno(1a)
$$

\noindent The components of the Euler equation:

$$
{\partial v_r \over \partial t} + v_r {\partial v_r \over \partial r} + {v_\phi \over r}
{\partial v_r \over \partial \phi} + v_z {\partial v_r \over \partial z} - {{v_\phi}^2 \over r}
= - {1 \over \rho} {\partial p \over \partial r} - {\partial g \over \partial r},
\eqno(1b)
$$

$$
{\partial v_\phi \over \partial t} + v_r {\partial v_\phi \over \partial r} + {v_\phi \over r}
{\partial v_\phi \over \partial \phi} + v_z {\partial v_\phi \over \partial z} + {1 \over r}
v_r v_\phi = - {1 \over \rho r} {\partial p \over \partial \phi},
\eqno(1c)
$$

\noindent and,

$$
{\partial v_z \over \partial t} + v_r {\partial v_z \over \partial r} + {v_\phi \over r}
{\partial v_z \over \partial \phi} + v_z {\partial v_z \over \partial z} = - {1 \over \rho}
{\partial p \over \partial z},
\eqno(1d)
$$

\noindent Energy equation:

$$
\rho v_r {\partial \epsilon \over \partial r} + \rho v_\phi {\partial \epsilon \over r \partial \phi}
+ \rho v_z {\partial \epsilon \over \partial z} + {\Lambda_{brems} + \Lambda_{sync}} =
- \rho {\partial \epsilon \over \partial t},
\eqno(1e)
$$

\noindent where, $g=-{GM \over {r-r_g}}$ is the Paczy{\'n}ski-Wiita pseudo-Newtonian potential, $G$ is the gravitational
constant, M is the mass of the black hole, $r_g={2GM \over c^2}$ is the Schwarzschild radius and
$c$ is the velocity of light.
The specific energy($\epsilon$) of the flow is
${1\over 2}(v^2_r + v^2_\phi + v^2_z) + {p\over \rho} {\gamma \over \gamma -1} + g$.
The expression for bremsstrahlung (Lang 1980) and synchrotron (Shapiro \& 
Teukolsky 1983) cooling
in an electron-proton plasma are,

$$
\Lambda_{brems}=1.43 \times 10^{-27} {\rho ^2 \over m_p ^2}
T^{1/2} g_f,
$$

\noindent and,

$$
\Lambda_{sync}= {16 \over 3}{e^2 \over c}({e \over {m_e c}})^2  {4\pi \over {\gamma(\gamma-1)}}
({1\over{m_ec^2}})^2 ({\mu m_p \over \gamma})^2 {1 \over m_p} { a^6 \rho ^2 },
$$

\noindent respectively, where $m_p$ is the mass of proton, $T$ is the temperature, $g_f$ is the Gaunt factor,
$e$ is the electron charge, $m_e$ is the electron mass,
$\gamma$ is the adiabatic index and $\mu=0.5$ for purely hydrogen gas.
Using the ideal gas equation, we get for the temperature $T$,

$$
T = {p \over \rho} \ {\mu m_p \over k},
$$

\noindent where, $k$ is the Boltzmann constant. The polytropic relation $p=K \rho ^ \gamma$ is
used to obtain the sound speed $a=\sqrt {\partial p \over
\partial \rho}=\sqrt{\gamma p \over \rho}$.

Frequencies that exist in the system are determined by a `dispersion relation'. Here in this case
we do not have an explicit equation for the dispersion relation
(See BH91 for an example of an explicit dispersion relation).
The $set$ of equations which are obtained
for infinitesimals can be considered as `dispersion relation'. The frequencies that solve this set
of equations are the eigen frequencies which exist in the physical system.
The variables of the perturbed flow are expressed as a sum of unperturbed value (which is the
steady state solution) and an
infinitesimal.
These variables are introduced in the flow equations and
simplified ignoring higher order terms. The equations are split into real
and imaginary parts to obtain following ten equations involving ten unknowns.

$
T_4 {v^{\prime}_{r_R}} = -\omega_R T_2 -\omega_I T_3
-v_{r_0} v^{\prime}_{r_0} v_{r_R} -v_{r_0} v^{\prime}_{\phi_0} v_{\phi_R}
-v_{r_0} v_{\phi_0} v^{\prime}_{\phi_R}
+{\gamma \over {\gamma-1}} ( v_{r_0} {\rho_R \over \rho^2} p^{\prime}_0 - 2v_{r_0}\rho_R p_0
{\rho^{\prime}_0 \over \rho^3_0} )+{\gamma \over {\gamma-1}} v_{r_0} {p_R \over \rho^2_0}
\rho^{\prime}_0
-v_{r_R} T_1
+{v_{\phi_0} \over r} m T_2
- {j g_f \over m^{3/2}_p} ({\mu \over k})^{1/2} \sqrt{p_0\rho_0} {1\over2}( {p_R \over p_0} + {\rho_R \over \rho_0} )
- {\gamma \over {\gamma-1}} T_5 {p_0 \over \rho^2_0} - {\gamma \over {\gamma-1}} v_{r_0}(-T_6
+ {\rho_R \over \rho^2_0} p^\prime_0),
$
$$
\eqno(2)
$$

$
T_4 {v^{\prime}_{r_I}} = -\omega_R T_3  -\omega_I T_2
-v_{r_0} v^{\prime}_{r_0} v_{r_I} -v_{r_0} v^{\prime}_{\phi_0} v_{\phi_I}
-v_{r_0} v_{\phi_0} v^{\prime}_{\phi_I}
+{\gamma \over {\gamma-1}} ( v_{r_0} {\rho_I \over \rho^2} p^{\prime}_0 - 2v_{r_0}\rho_I p_0
{\rho^{\prime}_0 \over \rho^3_0} )+{\gamma \over {\gamma-1}} v_{r_0} {p_I \over \rho^2_0}
\rho^{\prime}_0
-v_{r_I} T_1
+{v_{\phi_0} \over r} m T_3
- {j g_f \over m^{3/2}_p} ({\mu \over k})^{1/2} \sqrt{p_0\rho_0} {1\over2}( {p_I \over p_0} + {\rho_I \over
\rho_0} )
- {\gamma \over {\gamma-1}} T_7 {p_0 \over \rho^2_0} - {\gamma \over {\gamma-1}} v_{r_0}(-T_8
+ {\rho_I \over \rho^2_0} p^\prime_0),
$
$$
\eqno(3)
$$

$$
T_5 + \rho^\prime_R v_{r_0} + \rho_0 v^\prime_{r_R} = 0,
\eqno(4)
$$

$$
T_7 + \rho^\prime_I v_{r_0} + \rho_0 v^\prime_{r_I} = 0,
\eqno(5)
$$

$$
T_6 + v_{r_0} v^{\prime}_{r_R} = -{1 \over \rho_0} (p^{\prime}_R - {\rho_R \over \rho_0} p^\prime_0),
\eqno(6)
$$

$$
T_8 + v_{r_0} v^{\prime}_{r_I} = -{1 \over \rho_0} (p^{\prime}_I - {\rho_I \over \rho_0} p^\prime_0),
\eqno(7)
$$

$$
v_{\phi_R} \omega_I + v_{\phi_I} \omega_R + v_{r_0} v^{\prime}_{\phi_R} + v_{r_R} v^{\prime}_{\phi_
0} - {v_{\phi_0} \over r} v_{\phi_I} m + {1\over r} v_{r_0} v_{\phi_R} + {1\over r} v_{r_R}
v_{\phi_0} = {1\over \rho_0} {1\over r} p_I m,
\eqno(8)
$$

$$
v_{\phi_I} \omega_I - v_{\phi_R} \omega_R + v_{r_0} v^{\prime}_{\phi_I} + v_{r_I} v^{\prime}_{\phi_
0} + {v_{\phi_0} \over r} v_{\phi_R} m + {1\over r} v_{r_0} v_{\phi_I} + {1\over r} v_{r_I}
v_{\phi_0} = -{1\over \rho_0} {1\over r} p_R m,
\eqno(9)
$$

$$
v_{z_R} \omega_I + v_{z_I} \omega_R + v_{r_0} v^\prime_{z_R} - {v_{\phi_0} \over r} v_{z_I} m =
{1\over \rho_0} p_I k,
\eqno(10)
$$

\noindent and,

$$
v_{z_I} \omega_I - v_{z_R} \omega_R + v_{r_0} v^\prime_{z_I} + {v_{\phi_0} \over r} v_{z_R} m =
-{1\over \rho_0} p_R k,
\eqno(11)
$$

\noindent where,

$$
T_1 = v_{r_0} v^{\prime}_{r_0} + v_{\phi_0} v^{\prime}_{\phi_0} + {\gamma \over {\gamma-1}}
{\partial \over \partial r} ({p_0 \over \rho_0}) + g^{\prime},
$$

$$
T_2 = v_{r_0}v_{r_I} + v_{\phi_0}v_{\phi_I} + {\gamma \over ({\gamma-1})\rho_0} (-{\rho_I \over \rho_0} p_0 + p_I),
$$

$$
T_3 = v_{r_0}v_{r_R} + v_{\phi_0}v_{\phi_R} + {\gamma \over {(\gamma-1)}
\rho_0} (-{\rho_R \over \rho_0} p_0 + p_R),
$$

$$
T_4 = v_{r_0}^2 + {\gamma \over {\gamma-1}} {p_0 \over \rho_0} - {\gamma \over {\gamma-1}} v_{r_0}^2,
$$

$$
T_5 = \rho_R \omega_I + \rho_I \omega_R + {1 \over r} \rho_R v_{r_0} + \rho_R v^{\prime}_{r_0}
- {1 \over r} \rho_0 v_{\phi_I} m - {1 \over r} \rho_I m v_{\phi_0} - \rho_0 v_{z_I} k
+ {1 \over r} \rho_0 v_{r_R} + \rho^\prime_0 v_{r_R},
$$

$$
T_6 = v_{r_R} \omega_I + v_{r_I} \omega_R + v_{r_R} v^{\prime}_{r_0} - {v_{\phi_0} \over r}
v_{r_I} m - {2 v_{\phi_0} \over r} v_{\phi_R},
$$

$$
T_7 = \rho_I \omega_I - \rho_R \omega_R + {1 \over r} \rho_I v_{r_0} + \rho_I v^{\prime}_{r_0}
+ {1 \over r} \rho_0 v_{\phi_R} m + {1 \over r} \rho_R m v_{\phi_0} + \rho_0 v_{z_R} k
+ {1 \over r} \rho_0 v_{r_I} + \rho^\prime_0 v_{r_I},
$$

\noindent and,

$$
T_8 = v_{r_I} \omega_I - v_{r_R} \omega_R + v_{r_I} v^{\prime}_{r_0} + {v_{\phi_0} \over r}
v_{r_R} m - {2 v_{\phi_0} \over r} v_{\phi_I}.
$$

\noindent Here, $^\prime$ denotes the derivative with respect to $r$. Fourth-order Runge-Kutta
method (Press et al. 1992) is used to solve the equations (2-11).
The perturbations are introduced at the shock location and the equations
are integrated to obtain the functions, $\rho_R$, $\rho_I$, $v_{r_R}$, $v_{r_I}$, $v_{\phi_R}$, $v_{\phi_I}$, 
$v_{z_R}$, $v_{z_I}$, $p_R$ and $p_I$. The frequencies are changed over a grid.

\ \\
\noindent {\bf 4. Eigen frequencies}

\noindent The `dispersion relation' gives the frequencies that exist in the system. All the quantities oscillate
with that frequency but with different amplitudes.
As the shock location oscillates, the size of the post-shock region oscillates thereby intercepting
different amounts of soft X-ray flux, emitted by the pre-shock disc of a stellar mass black hole.
The fractional change in size of the post-shock region is large compared to that of the
pre-shock disc.
Hence QPOs are expected mainly in the hard X-ray flux.
For the case of super-massive black hole, the post-shock region radiates at a lower frequency.

We choose the mass of the stellar black hole to be
$14M_{\odot}$ and the value of $\gamma$ equal to 4/3, as distinct X-ray variability of GRS 1915+105 is interpreted
as due to instabilities in radiation pressure dominated disc (Greiner, Cuby and McCaughrean 2001).
The instability is decided by the
presence
of positive imaginary part of the perturbation frequency $\omega_I $.
In Manickam (2004, hereafter Paper I), the flow topologies and the parameter dependence on
the possibility of the formation of Rankine-Hugoniot shock was studied. It was 
mentioned that among
the different branches that are possible, for a given accretion rate and
specific angular momentum ($\lambda$), as a valid shock solution, stability and boundary conditions
should decide the uniqueness of the solution.
Here we consider the case, where accretion rate is equal to unity in Eddington units and specific angular
momentum is 1.8 in the units of $2GM/c$. The stability is analysed for bremsstrahlung cooling case and when both
bremsstrahlung and synchrotron
cooling are present.
The efficiency factor for synchrotron cooling process is chosen as $10^{-7}$ and $10^{-6}$, while that of
bremsstrahlung cooling process is chosen as unity.

The branches of flow topologies that are considered for stability analysis are shown in
Figs. 1(a-c). The stability of different branches when perturbations are introduced at
outer shock location are analysed.
The branches which form a shock are shown in $r_{shk}$ $vs$ $r_{cout}$ plot of Fig. 2.
We choose the supersonic branches corresponding to the outer critical point location at 23$r_g$,
25$r_g$ and
27$r_g$
for stability analysis.
The frequencies that exist in the system are obtained by numerically integrating the ten equations, (2-11),
using fourth-order Runge-Kutta method (Press et al. 1992). The following is the `flow chart' 
which is implemented in
the code.

    (i) the mass of the black hole is chosen to be $14M_{\odot}$, and $\gamma$ as 4/3

    (ii) set the efficiency factor of bremsstrahlung cooling ($zefac$) and synchrotron 
\indent cooling ($synfac$)

    (iii) obtain the flow topology (as described in Paper I) and find out the 
branches \indent that will
form a shock

    (iv) choose an outer supersonic branch and an inner subsonic branch that satisfies
\indent the shock conditions

    (v) choose the real and imaginary frequencies of perturbation, $\omega_R$ and $\omega_I$

    (vi) introduce perturbation at the shock location

    (v) integrate using fourth-order Runge-Kutta method to obtain the amplitude of 
\indent the 
perturbation
in the subsonic branch
   (supersonic branches are not considered \indent for stability analysis as
the perturbations
would get advected with the flow in \indent infall time scale)

    (vi) frequencies are changed over a grid

    (vii) if a positive imaginary frequency exists, then the accretion disc is unstable

    (viii) if perturbation amplitude blows up for even one of the flow variables, then \indent the flow
is not amenable to a linear analysis

\ \\
\noindent {\bf 5. Results}

\noindent The Fig. 3(a-b), Fig. 4(a-b) and Fig. 5(a-b) show the frequencies
that exist in the system.
  These are plots of $({\omega_I / 2 \pi})$ $vs$ $({\omega_R / 2\pi})$.
Hence values in the axis are in hertz. It is of the order of 10 Hz which
is same as that observed in galactic black hole candidate GRS 1915+105.
 For the case of synchrotron efficiency factor of $10^{-7}$, there is not much effect
in the frequency modes that exist in the system,
due to
addition of synchrotron cooling process.

But when the efficiency factor is increased to $10^{-6}$, the flow topology changes drastically
(Fig. 1c) and the flow is
free of shock. This is in line with the observation of CLS85, where shock disappears
if the magnetic field is above a certain value.
 Stability analysis with more realistic models could explain the observed QPOs better.
For instance, the effect of magnetic field on the dynamics of the flow is not considered.
 Magnetic field seems to be present in GRS 1915+105 and it plays an important role (Nandi, Chakrabarti,
 Vadawale \& Rao  2001; Vadawale, Rao \& 
 Chakrabarti 2001).

\ \\
\noindent {\bf 6. Discussion and Conclusions}

\noindent Not all the frequency
ranges fall within the linear analysis domain, and the failure of linear analysis does
not imply instability or stability. Like the accreting fluid has infinite degrees of
freedom, there are many cases which can be considered for linear analysis. Here only a
few representative cases are illustrated. These results could be useful to test
a time-dependent numerical code which could handle non-linear perturbation analysis.

\newpage
\hskip -2cm
\begin{picture}(4,85){
\epsfxsize=12cm
\epsfysize=18cm
\epsfbox{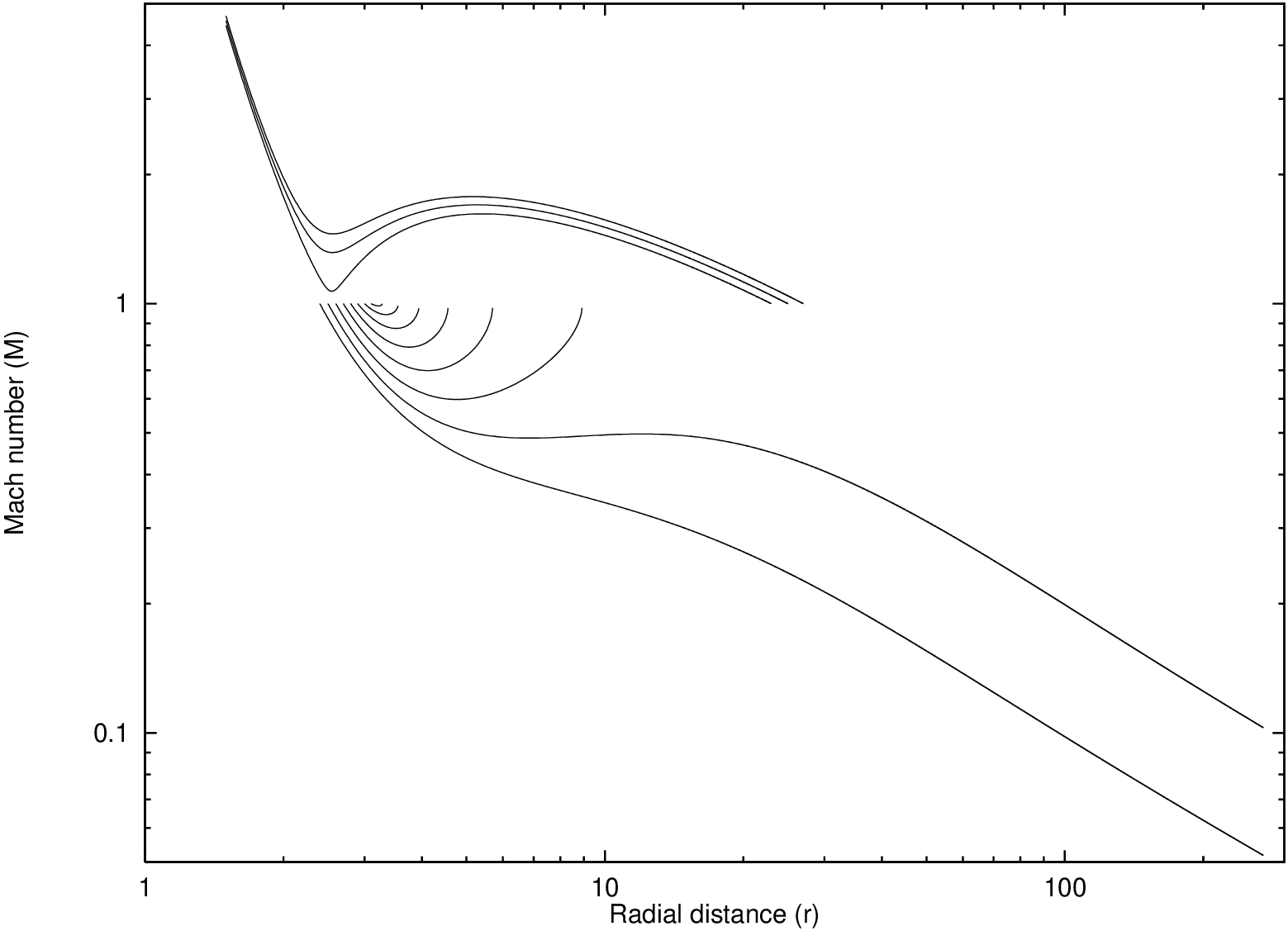}
}
\end{picture}
\vspace{15cm}
\\ Fig. 1a: The branches of the flow topology considered for stability analysis.
Supersonic branches have critical point locations at 23$r_g$, 25$r_g$ and 27$r_g$.
Efficiency of bremsstrahlung cooling is unity and there is no synchrotron cooling.
Mass of stellar black hole is $14M_{\odot}$
and $\gamma=4/3$.
Accretion rate is 1.0 in Eddington units
and $\lambda$=1.8 in units of $2GM/c$.

\newpage
\hskip -2cm
\begin{picture}(4,85){
\epsfxsize=12cm
\epsfysize=18cm
\epsfbox{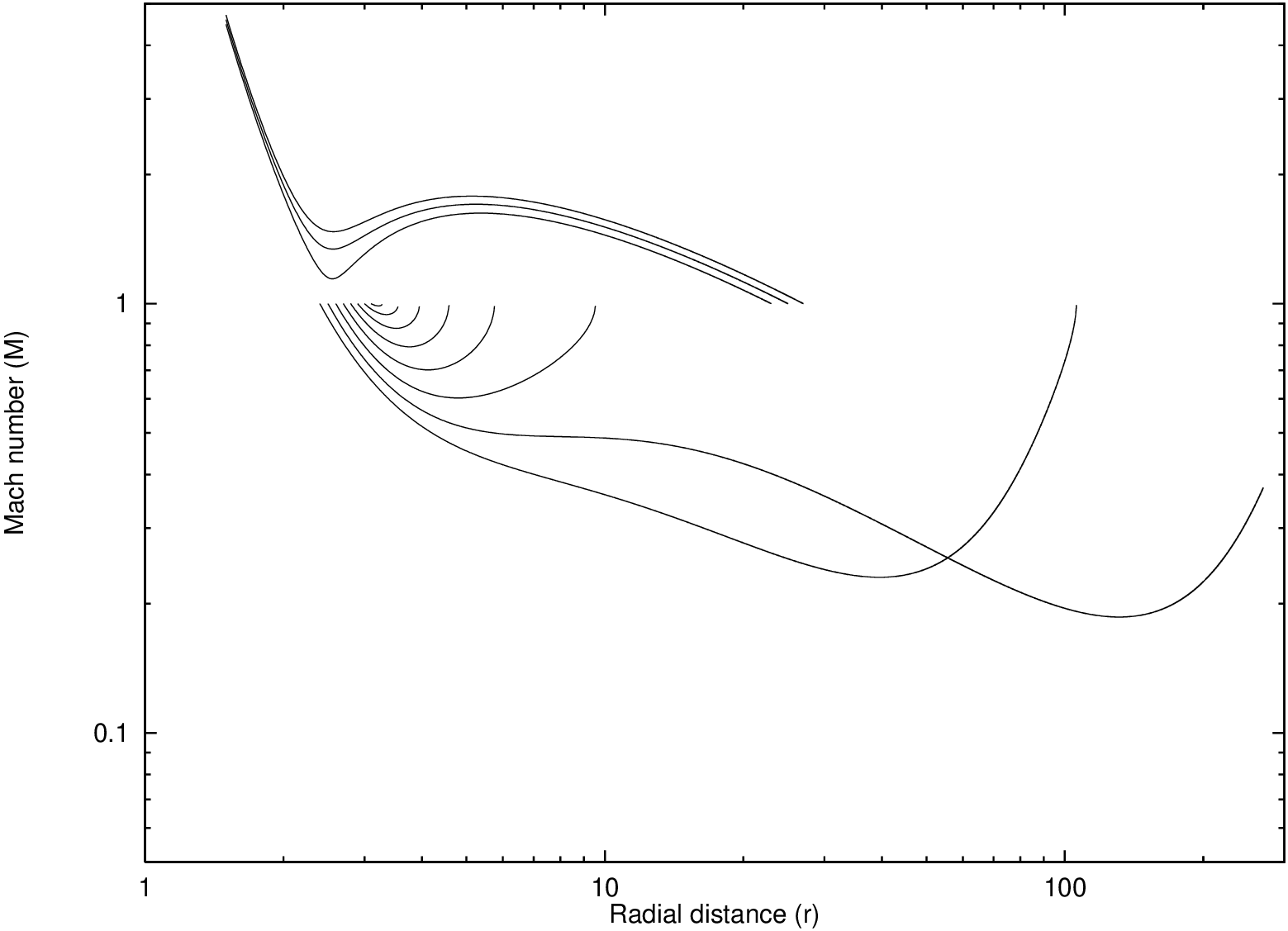}
}
\end{picture}
\vspace{15cm}
\\ Fig. 1b: Same as Fig. 1a but synchrotron cooling of efficiency factor $10^{-7}$
is included.

\newpage
\hskip -2cm
\begin{picture}(4,85){
\epsfxsize=12cm
\epsfysize=18cm
\epsfbox{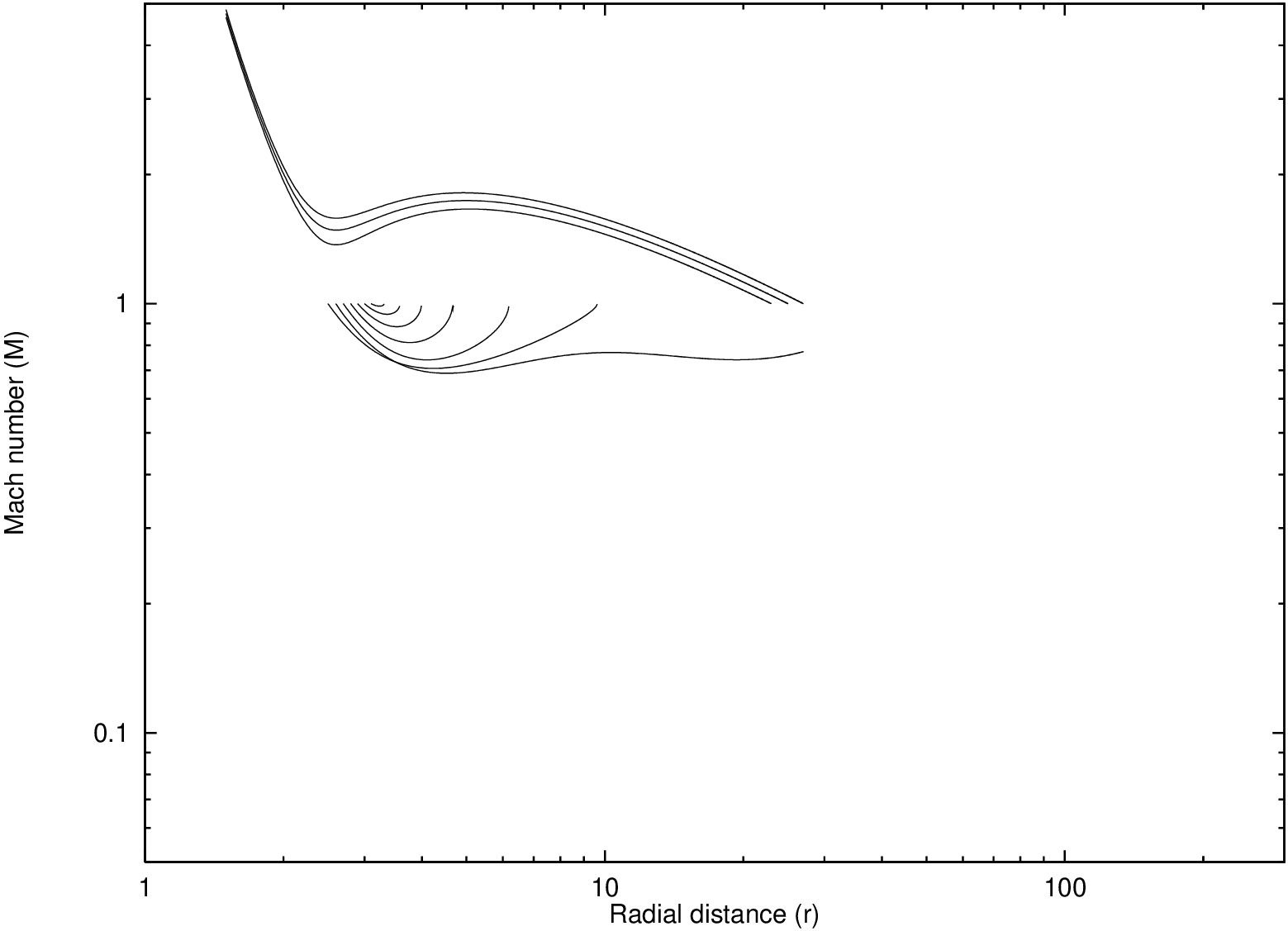}
}
\end{picture}
\vspace{15cm}
\\ Fig. 1c: Same as Fig. 1a but synchrotron cooling of efficiency factor $10^{-6}$
is included.

\newpage
\begin{picture}(4,365){
\epsfxsize=12cm
\epsfysize=18cm
\epsfbox{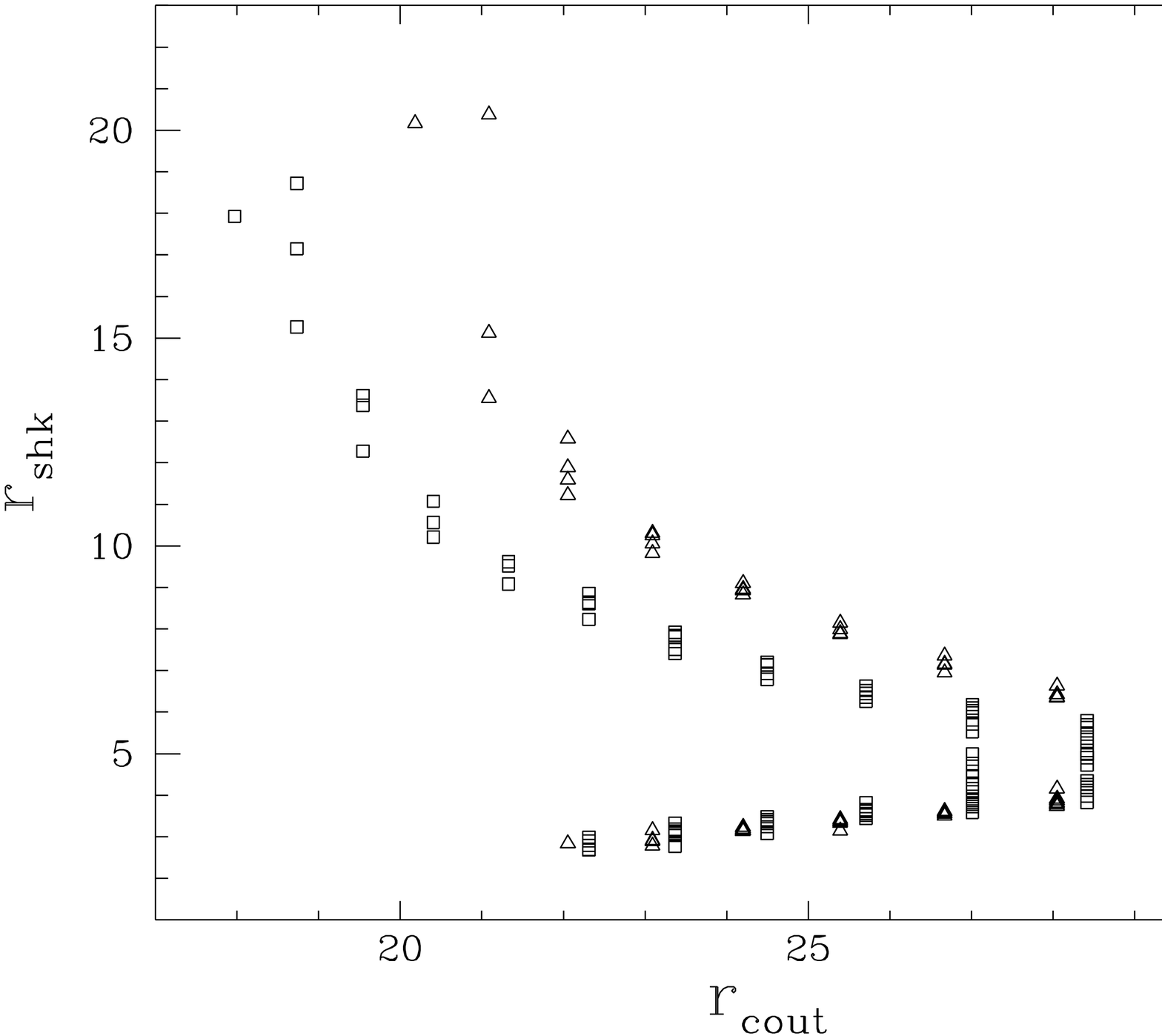}
}
\end{picture}
\vspace{2cm}
\\ Fig. 2: Outer critical point location ($r_{cout}$) of the supersonic branches that forms shock
at location $r_{shk}$.
The left set of points correspond to bremsstrahlung plus $10^{-7}$ efficient synchrotron
cooling process and right set is for bremsstrahlung cooling. The parameters are
same as that of Figs. 1(a-b).

\newpage
\begin{picture}(4,250){
\epsfxsize=10.5cm
\epsfysize=12cm
\epsfbox{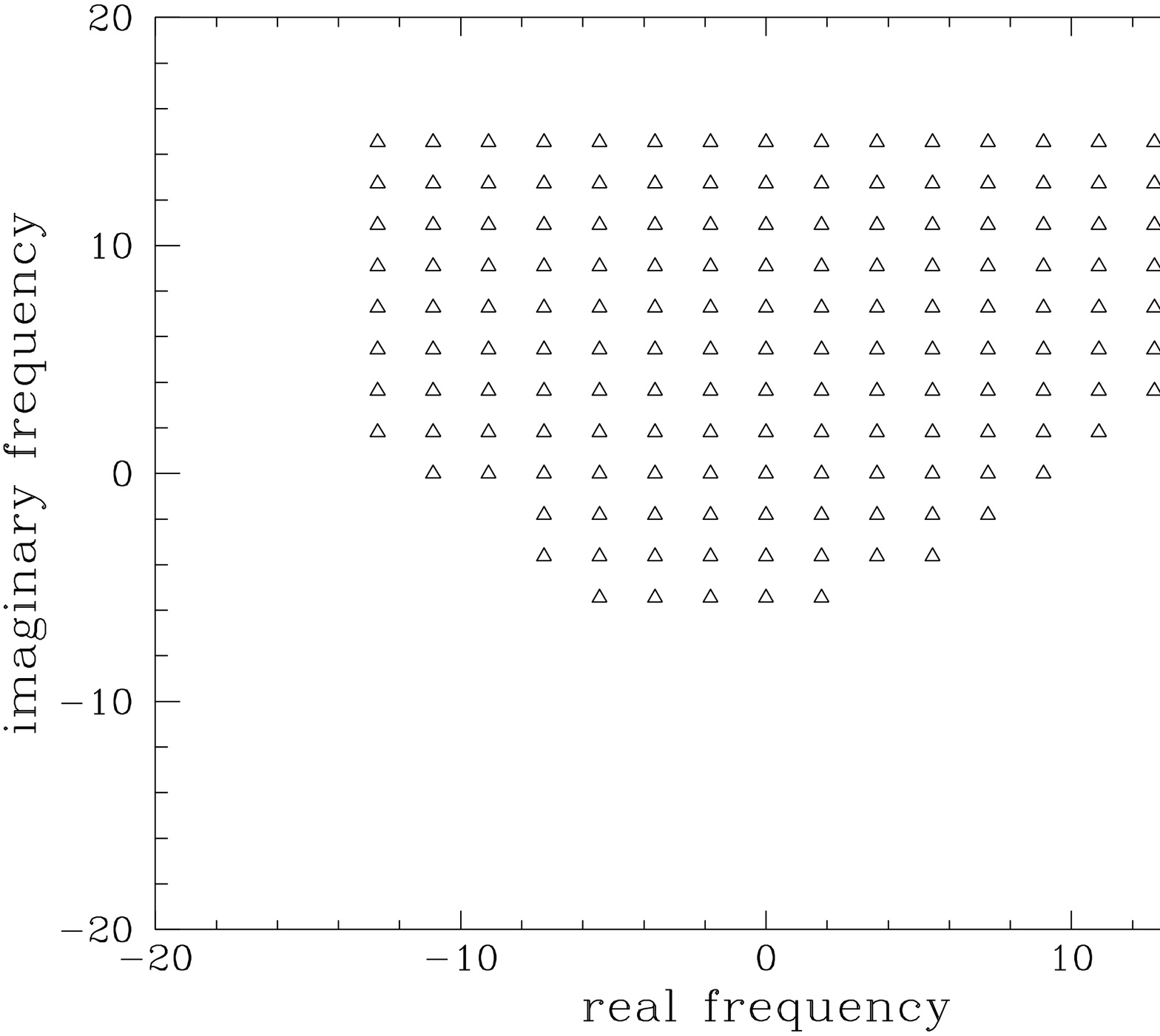} }
\end{picture}

\begin{picture}(4,250){
\epsfxsize=10.5cm
\epsfysize=12cm
\epsfbox{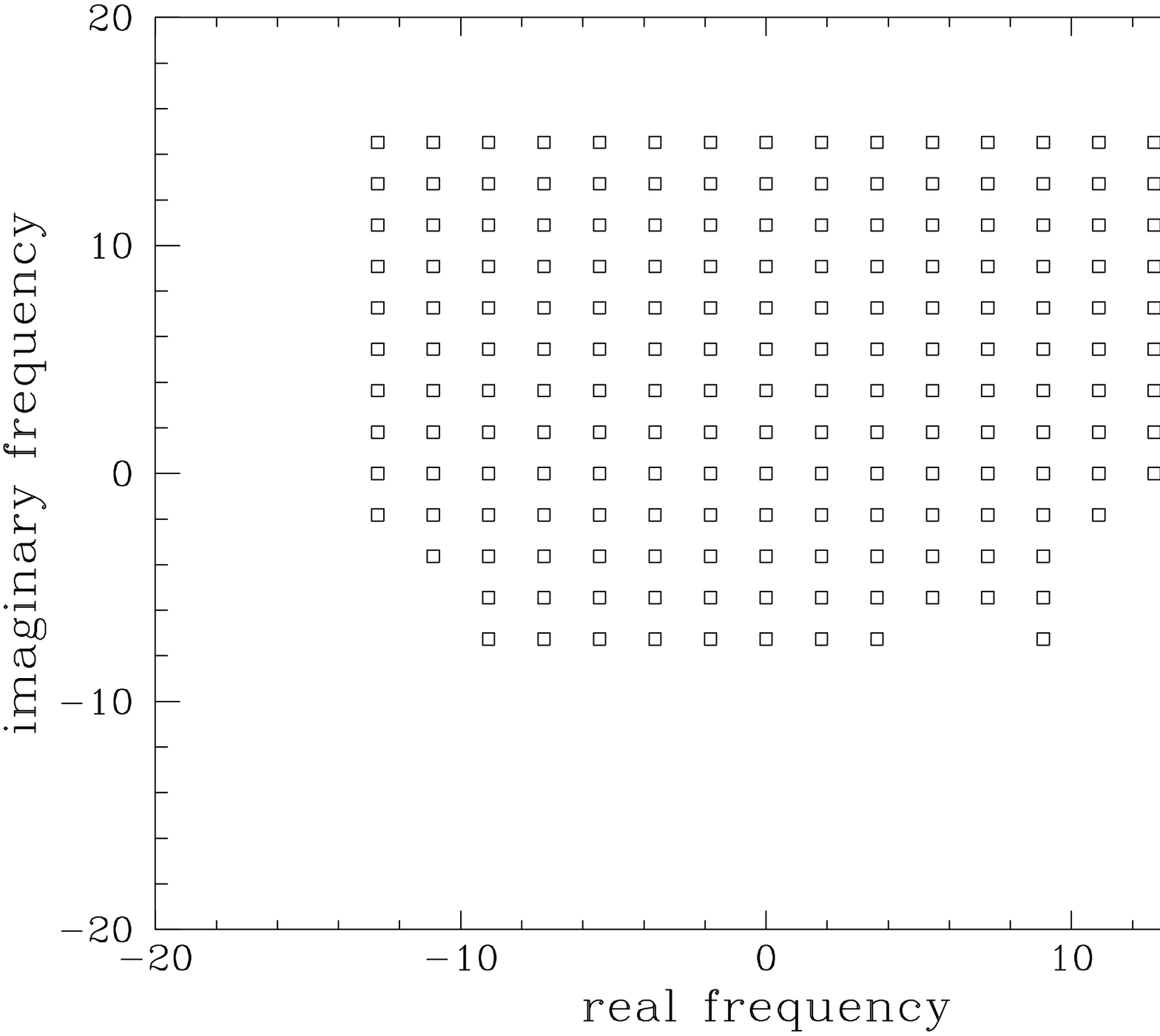} }
\end{picture}
\vspace{1.2cm}
\\ Fig. 3(a-b): This shows the frequency modes that exist in the accretion disc
for outer supersonic branch of critical point location at $23r_g$. The positive imaginary frequencies
are the unstable modes. Bremsstrahlung cooling case is shown
in (a) and bremsstrahlung plus $10^{-7}$ efficient synchrotron cooling is shown in (b).

\newpage
\begin{picture}(4,250){
\epsfxsize=10.5cm
\epsfysize=12cm
\epsfbox{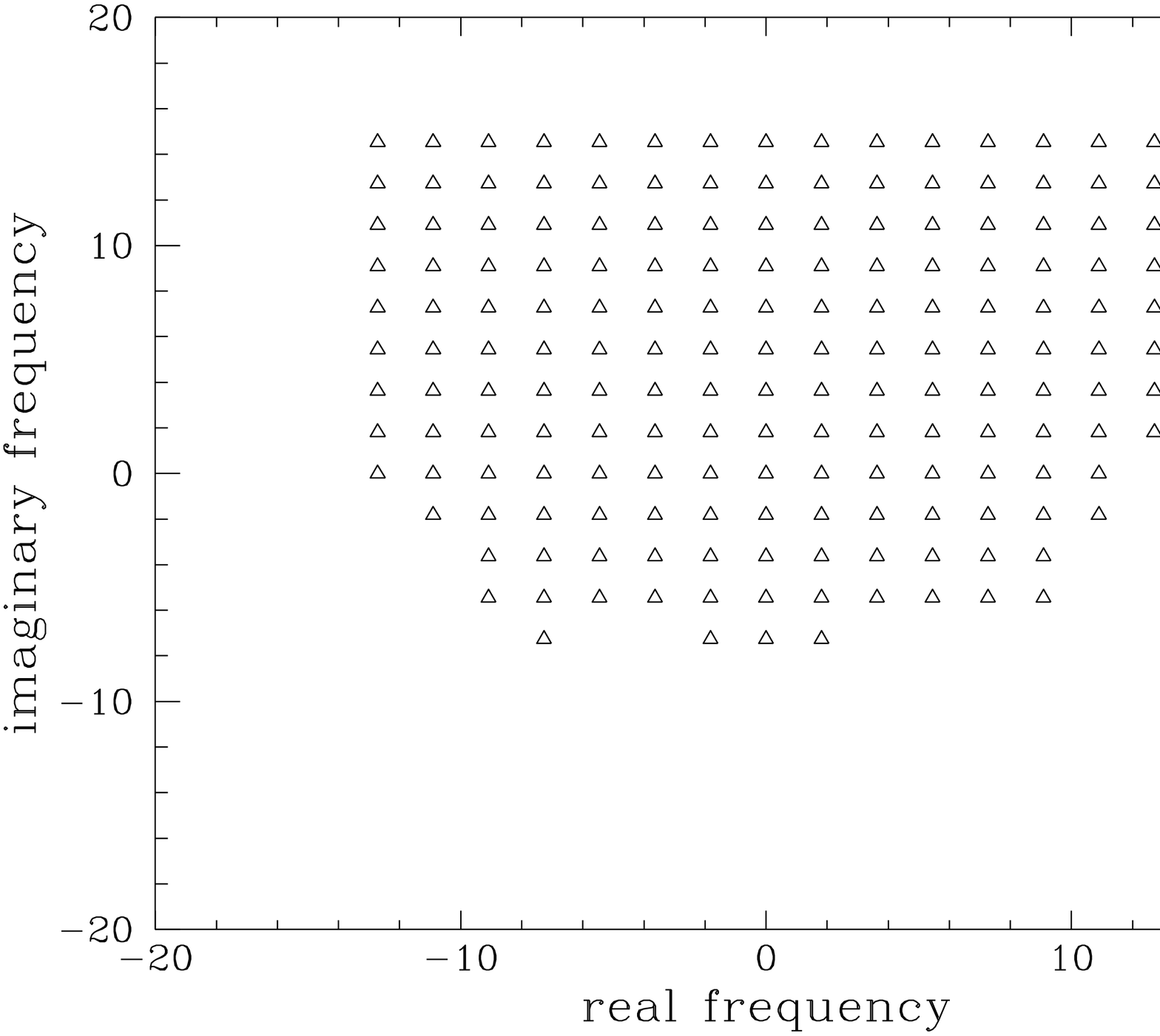} }
\end{picture}

\begin{picture}(4,250){
\epsfxsize=10.5cm
\epsfysize=12cm
\epsfbox{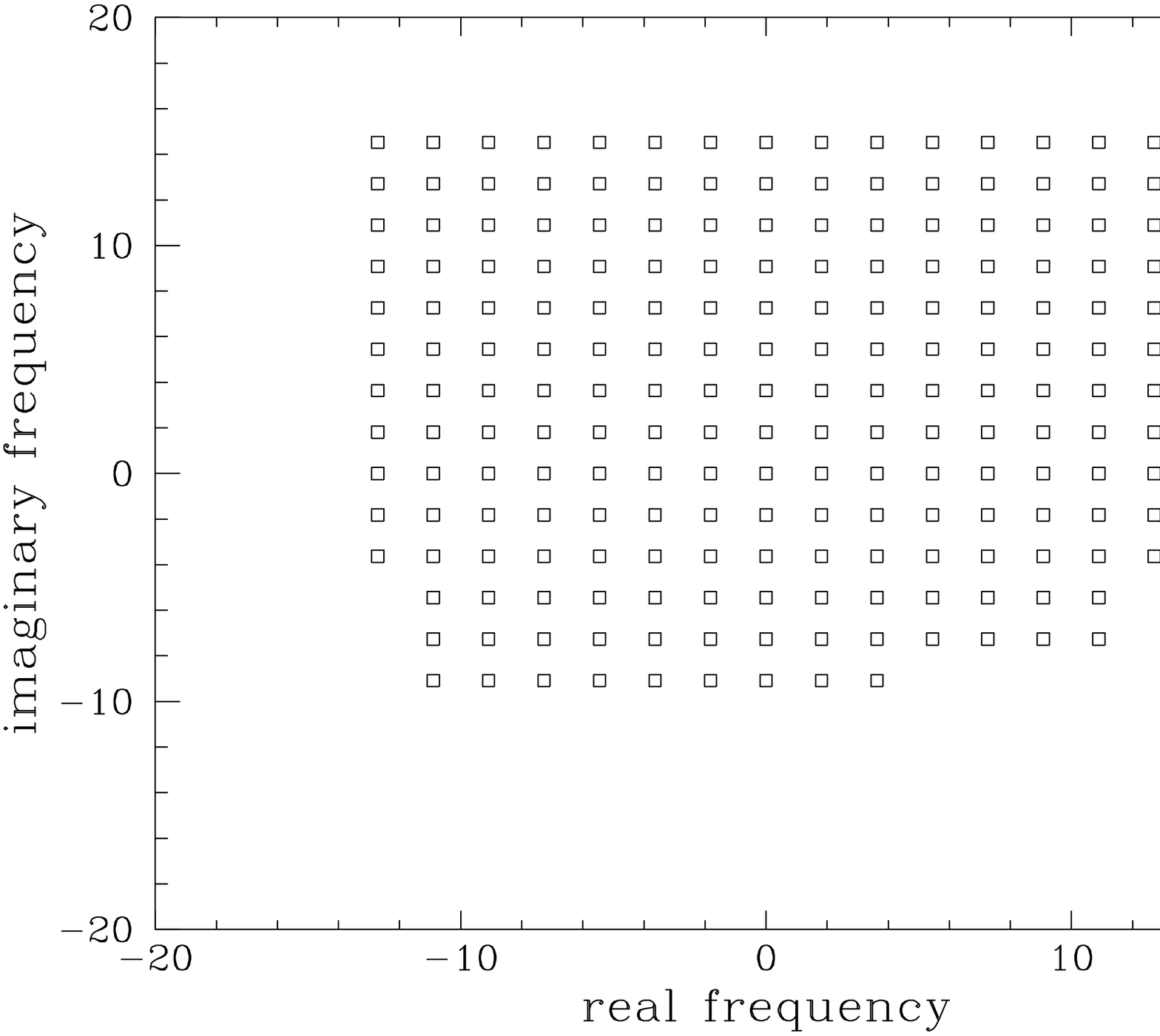} }
\end{picture}
\vspace{1.2cm}
\\ Fig. 4(a-b): Same as Fig. 3(a-b) but outer supersonic branch has critical point location at
$25r_g$.

\newpage
\begin{picture}(4,250){
\epsfxsize=10.5cm
\epsfysize=12cm
\epsfbox{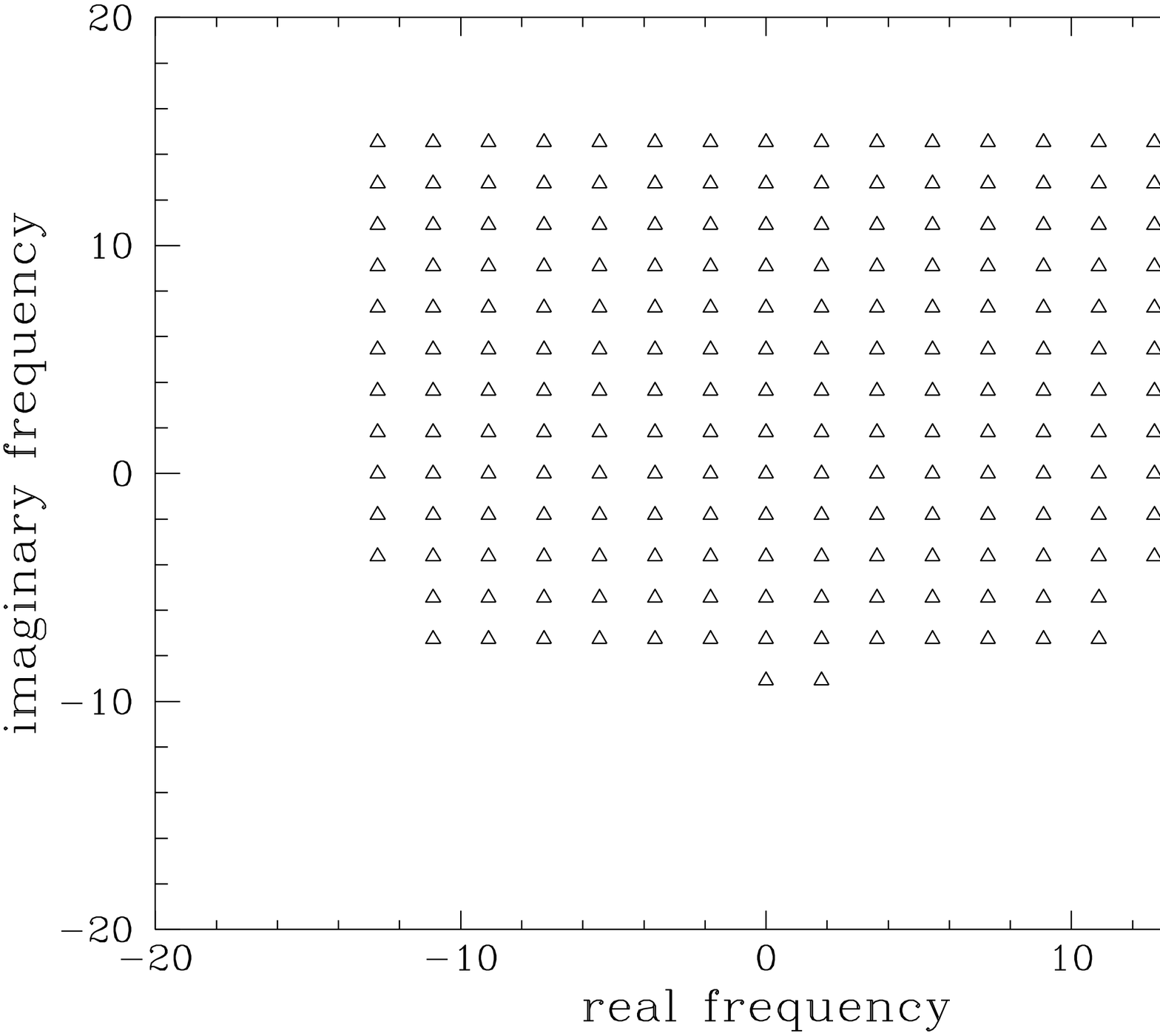} }
\end{picture}

\begin{picture}(4,250){
\epsfxsize=10.5cm
\epsfysize=12cm
\epsfbox{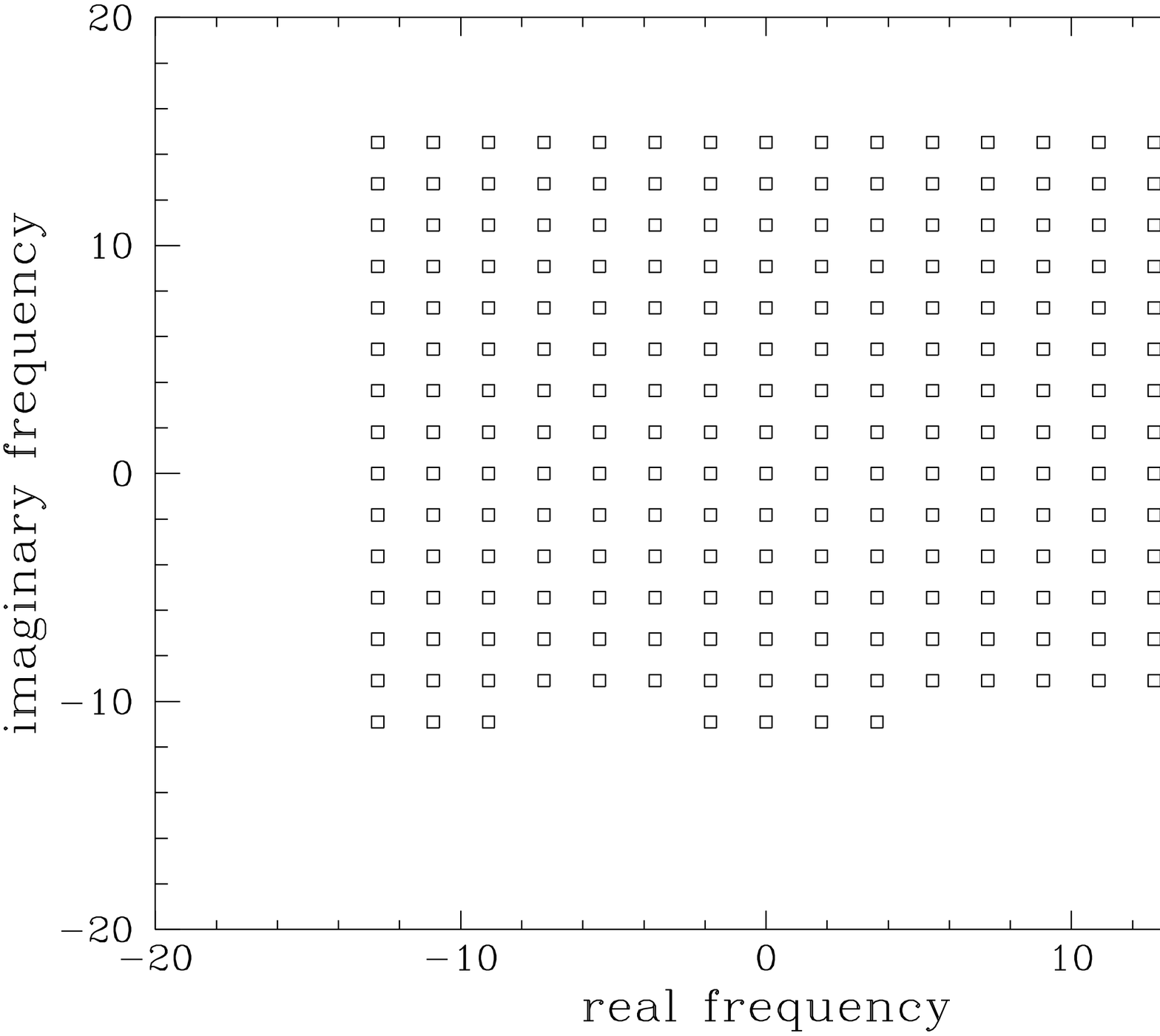} }
\end{picture}
\vspace{1.2cm}
\\ Fig. 5(a-b): Same as Fig. 3(a-b) but outer supersonic branch has critical point location at
$27r_g$.

\newpage 
\noindent {\bf {References}}

\ \\
\noindent Balbus, S. A. and  Hawley, J. F. 1991, ApJ, 376, 214, (BH91)

\noindent Bondi, H. 1952, MNRAS, 112, 195

\noindent Chakrabarti, S. K. and Manickam, S. G. 2000, ApJ, 531, L41

\noindent Chandrasekhar, S. 1961, Hydrodynamic and Hydromagnetic stability, Dover 
\\ \indent Publications, Inc., New York

\noindent Chanmugam, G., Langer, S. H. and Shaviv, G. 1985, 299, L87, (CLS85)

\noindent Greiner, J., Cuby, J. G., and McCaughrean, M. J. 2001, Nature, 414, 522

\noindent Lang, K. R. 1980, Astrophysical Formulae, second edition, Springer-Verlag

\noindent Langer, S. H., Chanmugam, C. and Shaviv, G. 1982, ApJ, 258, L289, (LCS82)

\noindent Manickam, S. G. 2004, astro-ph/0410713, (Paper I)

\noindent Molteni, D., Sponholz, H. and Chakrabarti, S. K. 1996, ApJ, 457, 805

\noindent Nandi, A., Chakrabarti, S. K.,
 Vadawale, S. V. and  Rao, A. R. 2001, A\&A, 380, 245

\noindent Narayan, R. and Yi, I. 1994 ApJ, 428, L13

\noindent Novikov, I. D. and Thorne, K. S. 1973, in Black Holes, ed. B. S.
     De Witt and 
\\ \indent C. De Witt (New York: Gordon \& Breach), 343

\noindent Paczy{\'n}ski, B. 1998, Acta Astronomica, 48, 667

\noindent Paczy{\'n}ski, B. and Wiita, P. J. 1980, A\&A, 88, 23

\noindent Press, W. H., Teukolsky, S. A., Vellerling, W. T. and Flannery, B. P. 1992,
Numerical 
\\ \indent Recipes in Fortran - The Art of Scientific Computing, second edition,
Cambridge 
\\ \indent University Press

\noindent Shakura, N. I. and Sunyaev, R. A. 1973, A\&A, 24, 337

\noindent Shapiro, S. L., and Teukolsky, S. A. 1983, Black Holes, White Dwarfs,
and Neutron 
\\ \indent Stars, New York; Wiley

\noindent Vadawale, S. V., Rao, A. R. and
 Chakrabarti, S. K. 2001, A\&A, 372, 793

\end{document}